\documentclass[twocolumn,showpacs,unsortedaddress,prc,floatfix,amsmath,amssymb,aps]{revtex4-1}
\usepackage{dcolumn,pstricks,bm,booktabs}
\usepackage{graphicx}
\newcommand{\tsups}[1]{\textsuperscript{#1}}  

\newcommand{\trinat}{{\scshape Trinat}}
\newcommand{\triumf}{{\scshape Triumf}}

\begin{document}

\title{Precision half-life measurement of the \boldmath$\beta^+$ 
  decay of \tsups{37}K}
\date{\today}

\author{P.D.\ Shidling$^1$}
\email{pshidling@comp.tamu.edu}
\author{D.\ Melconian$^{1,2}$}
\author{S.\ Behling$^{1,3}$}
\author{B.\ Fenker$^{1,2}$}
\author{J.C.\ Hardy$^{1,2}$}
\author{V.E.\ Iacob$^{1}$}
\author{E.\ McCleskey$^{1,2}$}
\author{M.\ McCleskey$^{1,2}$}
\author{M.\ Mehlman$^{1,2}$}
\author{H.I.\ Park$^{1,2}$ and B.T.\ Roeder$^{1}$}
\affiliation{$^1$ Cyclcotron Institute, Texas A\&M University, College Station, Texas, 77843-3366}
\affiliation{$^2$ Department of Physics, Texas A\&M University,College Station, Texas, 77843-4242}
\affiliation{$^3$ Department of Chemistry, Texas A\&M University,College Station, Texas, 77843-302}\def\andname{}

\begin{abstract}
  The half-life of \tsups{37}K has been measured to be $1.23651(94)~\mathrm{s}$,
  a value nearly an order of magnitude more precise than the best previously 
  reported.  The $\beta^+$ decay of \tsups{37}K occurs mainly via a 
  superallowed branch to the ground-state of its $T=1/2$ mirror, \tsups{37}Ar. 
  This transition has been used recently, together with similar transitions 
  from four other nuclei, as an alternative test of CVC and method for 
  determining $V_{ud}$, but the precision of its $ft$ value was limited by 
  the relatively large half-life uncertainty.  Our result corrects that 
  situation.  Another motivation for improving the $ft$ value was to determine 
  the standard-model prediction for the $\beta$-decay correlation parameters, 
  which will be compared to those currently being measured by the 
  \trinat{} collaboration at \triumf. The new $ft$ value, $4605(8)~\mathrm{s}$, 
  is now limited in precision by the $97.99(14)\%$ ground-state branching 
  ratio.
\end{abstract}

\date{\today}

\maketitle
\section{Introduction}
Many precision measurements of the correlation parameters and $ft$ values of 
$\beta$-decaying nuclei have been used to help form our understanding of the 
fundamental symmetries of the weak interaction.  Experiments of this kind 
continue to be performed to search for physics beyond the standard model.
For example, the measured $ft$ values for superallowed $J^\pi=0^+\!\!\rightarrow
0^+$ pure Fermi transitions, have been used to verify the conserved vector 
current (CVC) hypothesis to one part in $10^4$; to determine most precisely 
the value of $V_{ud}$, the up-down element of the Cabbibo-Kobayashi-Maskawa 
(CKM) matrix; and to set limits on scalar and right-handed 
currents~\cite{HardyPRL:05,Hardy:05,Hardy:09}.
    
Transitions between isospin $T=1/2$ doublets in mirror nuclei are 
also useful weak-interaction probes because there are relatively few 
corrections required to describe their decay.  The most thoroughly studied 
and theoretically cleanest example of such a transition is the decay of the
neutron~\cite{mendenhallPRC2013,plasterPRC2012,PERKEO}, but a number of other 
cases have attracted attention recently: the decays of \tsups{19}Ne, 
\tsups{21}Na, \tsups{29}P, \tsups{35}Ar and the subject of the present 
work, \tsups{37}K.  Naviliat-Cuncic and Severijns~\cite{NavPRL09} used the 
$ft$ values for these five transitions to determine $V_{ud}$, though less 
precisely than has been achieved from the $0^+\!\!\rightarrow0^+$ transitions. 
With more precise measurements, however, the $T=1/2$ result for $V_{ud}$ could 
be improved enough to provide a valuable cross-check on the result from the 
$0^+\!\!\rightarrow0^+$ transitions.  Our measurement is a first step on this 
path.

For mixed Fermi/Gamow-Teller transitions, such as the decay of \tsups{37}K to 
the ground state of \tsups{37}Ar, the measured $ft$ value can be related to 
$V_{ud}$ via the corrected $\mathcal{F}t$ value, which incorporates calculated 
corrections for radiative and isospin-symmetry-breaking effects, as 
follows~\cite{SevPRC08}: 
\begin{align}
  \mathcal{F}t &= f_V t(1+\delta'_R)(1+\delta_\mathrm{NS}^V-\delta_C^V) \nonumber\\
  &=\frac{K}{G_F^2|V_{ud}|^2\,C_V^2|M_F^0|^2\big(1+\Delta_R^V\big)
    \Big[1+\Big(\frac{f_A}{f_V}\Big)\rho^2\Big]}\label{eq:Ft}
\end{align}
where $K/(\hbar c)^6=2\pi^3\ln{2}\hbar/(m_ec^2)^5=8120.278(4)\times
10^{-10}~\mathrm{GeV}^{-4}\mathrm{s}$, the Fermi constant $G_F/(\hbar c)^3=
1.16637(1)\times10^{-5}~\mathrm{GeV}^{-2}$, and the vector coupling constant 
$C_V=1$.  The parameters $\delta'_R$, $\delta_C^V$, $\delta_\mathrm{NS}^V$ and 
$\Delta_R^V$ are the usual radiative and isospin-symmetry-breaking correction 
terms calculated for the vector current \cite{Hardy:09}. The statistical rate
functions for vector and axial-vector currents are denoted $f_V$ and $f_A$.  
In these mirror nuclei their ratio $f_A/f_V$ generally is within a few percent 
of unity according to shell model calculations~\cite{SevPRC08}.  

The Fermi matrix element in the limit of strict isospin-symmetry is 
\begin{align}
  M_F^0=\big[T(T+1)-T_Z(T_Z\pm1)\big]^{1/2}
\end{align}
for a $\beta^\pm$ transition from a nucleus having $Z$ protons, $N$ neutrons, 
and isospin $T_Z=\frac{1}{2}(N-Z)$.  It takes the value 1 for $T$=1/2 mirror 
transitions.

Finally, the Gamow-Teller to Fermi mixing ratio $\rho$ is given by
\begin{align}
  \rho=\frac{C_AM_{GT}^0}{C_VM_F^0}\left[
    \frac{(1+\delta^A_\mathrm{NS}-\delta^A_C)(1+\Delta^A_R)}
    {(1+\delta^V_\mathrm{NS}-\delta^V_C)(1+\Delta^V_R)}
  \right]^{1/2},\label{eq:rho}
\end{align}
where $M_{GT}^0$ is the Gamow-Teller matrix element, $C_A$ is the axial-vector
coupling constant, and the correction terms with $A$ superscripts are 
equivalent to those with $V$ superscripts but must be evaluated for the 
axial-vector component.  Since, in general, $C_AM_{GT}^0$ cannot be precisely 
calculated, $\rho$ must be taken from experiment.  This also makes it 
unnecessary to calculate the axial-vector correction terms. 
 
The decay of \tsups{37}K was one of the five cases used in 
Ref.~\cite{NavPRL09} to determine $V_{ud}$.  The relatively large uncertainty 
of its contribution was overwhelmingly dominated by the precision of $\rho$, 
which was obtained from a single $\pm3\%$ measurement of the neutrino 
asymmetry parameter, $B_\nu$~\cite{MelconianPLB2007}.  The $\mathcal{F}t$ 
value is known much better, to $\pm0.6\%$~\cite{SevPRC08}, with the largest 
contribution to its uncertainty being the lifetime of the decay.  A survey 
of mirror transitions and the theoretical corrections to their $f_Vt$ values 
was made in Ref.~\cite{SevPRC08}, in which three \tsups{37}K half-life 
measurements~\cite{Sc58, Ka64, Az77} were averaged to arrive at 
$t_{1/2}=1.2248(73)~\mathrm{s}$.  This $\pm0.60\%$ uncertainty is much larger 
than the $\pm0.14\%$ and $\pm0.04\%$ uncertainties on the other two 
contributors to $f_Vt$ for \tsups{37}K: the branching ratio and $f_V$ value,
respectively.

The aim of the present work is to improve the lifetime of \tsups{37}K so that 
it no longer dominates the uncertainty in the deduced $\mathcal{F}t$ value.  
A more precise $\mathcal{F}t$ value on its own will not improve the value 
of $V_{ud}$ obtained from \tsups{37}K decay.  For that purpose, an improved 
measurement of $\rho$ via a correlation parameter such as $B_\nu$ or $A_\beta$ is 
necessary.  

\section{Experiment}
A high purity radioactive beam is essential for performing a 
precision half-life measurement. We have measured the half-life of \tsups{37}K 
at the Cyclotron Institute of Texas A\&M University using a 
$4\pi$ continuous-gas-flow proportional counter and a fast 
tape-transport system. We produced \tsups{37}K via an 
inverse kinematics reaction\;(fusion-evaporation reaction) by accelerating a primary beam of \tsups{38}Ar 
to $29~\mathrm{MeV}/\mathrm{u}$ in the K500 superconducting cyclotron, and 
impinging it on a $1.6$-atm\ H$_2$ gas target cooled to liquid 
nitrogen temperature.  The forward focused reaction products were analyzed 
using the Momentum Achromat Recoil Separator 
(MARS)~\cite{mars,tamu-beams+MARS}. Selection and focusing of the desired 
ions in MARS was done according to their $m/q$ value. At the focal plane of 
MARS, a 16-strip position-sensitive detector of $300~\mu\mathrm{m}$ thickness 
with an active area of $5\times5~\mathrm{cm}^2$ could be inserted to 
detect the secondary reaction products, which were identified according to 
their position and energy loss in the strip detector. The position of the 
reaction products on the strip detector was related with its $m/q$ ratio. 
This detector was initially inserted to help us tune the spectrometer 
and was then removed before the measurement began. During the measurement, 
it was reinserted once a day to ensure that no changes had occurred. 
In this way we determined that the 22.7-$\mathrm{MeV/u}$ \tsups{37}K 
beam passing through the spectrometer extraction slits was 
$98.4\%$ pure, the remaining contaminants being \tsups{35}Ar, \tsups{34}Cl 
and \tsups{33}Cl.

This separated beam then exited the vacuum system 
through a $51~\mu\mathrm{m}$ thick Kapton foil. Immediately following this 
foil was a $0.3~\mathrm{mm}$-thick BC-404 plastic scintillator, which counted 
the number of ions.  The beam then passed through a stack of Al degraders, 
and finally was implanted into the 76-$\mu\mathrm{m}$-thick aluminized Mylar 
of the fast-tape transport system.  The degrader thickness was adjusted so
as to place the implanted \tsups{37}K at a depth mid-way through the tape.  At
this setting ($51\,\mu$m of Al), the combination of electromagnetic separation 
in MARS and range selection in the degraders resulted in the \tsups{37}K beam 
reaching the tape being $>98.8\%$ pure.

For part of the measurement we used two different thicknesses of degrader: 
$70\,\mu$m, which resulted in the \tsups{37}K being placed at the front of 
the tape; and $13\,\mu$m, which put it at the back of the tape.  The former 
led to a higher proportion of impurities; the latter to a much lower 
proportion. In both cases the intensity of \tsups{37}K was reduced.  In the 
best case, with the \tsups{37}K deposited near the back of the tape, the 
sample purity was $>99.7\%$.

A sample of \tsups{37}K was collected by implanting the beam into a section 
of aluminized Mylar tape for a time interval of $0.5~\mathrm{s}$. The beam 
was then turned off and the tape-transport system moved the sample in 
$170~\mathrm{ms}$ to a well-shielded location, stopping it in the center of 
a $4\pi$ proportional gas counter. There, the $\beta$ particles from the 
decay were counted for $\approx20$ half-lives ($25~\mathrm{s}$).  This
cycle was repeated continuously, with each sample being collected on
a fresh region of tape, until the desired statistics had been accumulated.

The $4\pi$ gas counter and the data acquisition system used in the present 
measurement were similar to those originally described in 
Ref.~\cite{koslowskyNIMA1997}.  The current system has also been described 
in detail in reports of previous half-life measurements of superallowed 
$\beta^{+}$ emitters performed at Texas A\&M University (see, for example, 
\cite{iacobPRC2006,iacobPRC2010}).  Briefly, the preamplified signals from 
the gas counter are passed to a fast-filter amplifier with a high gain 
($\times500$). At this high gain many of the pulses would saturate the 
amplifier so, to ensure that the amplifier recovers quickly, large pulses 
are clipped with a Schottky diode inserted after the first stage of 
amplification. The amplified and clipped pulses are then sent to a 
discriminator with very low threshold ($150-250$~mV).

Since deadtime is a serious concern for half-life measurements, the 
discriminator signals were split and sent to two fixed-width nonretriggering 
and nonextending gate generators, which established different dominant 
deadtimes in the two separate streams, both of which were multiscaled into 
500-channel time spectra.  

The total measurement was split into 13 runs with different experimental 
conditions: three discriminator threshold settings (150, 200 and 250\,mV), 
two combinations of dominant deadtimes (4 and 6\,$\mu$s; 3 and 8\,$\mu$s), 
three detector bias voltages within the detector's plateau region (2600, 2650 
and 2700\,V) and three degrader thicknesses (13, 51 and 70\,$\mu$m).  The 
initial activity of \tsups{37}K for each cycle was in the range of 
$4000-7000~\mathrm{cps}$. All runs were composed of $100-300$~cycles, which 
yielded a total of $4\times10^6$ counts in each. In addition, a background 
measurement was performed for which all conditions were the same as for 
normal data taking except that the tape motion was disabled. The background 
rate was found to be four orders of magnitude lower than the initial count 
rate for each collected sample.

\begin{figure}\centering
  \includegraphics[angle=90,width=0.485\textwidth]{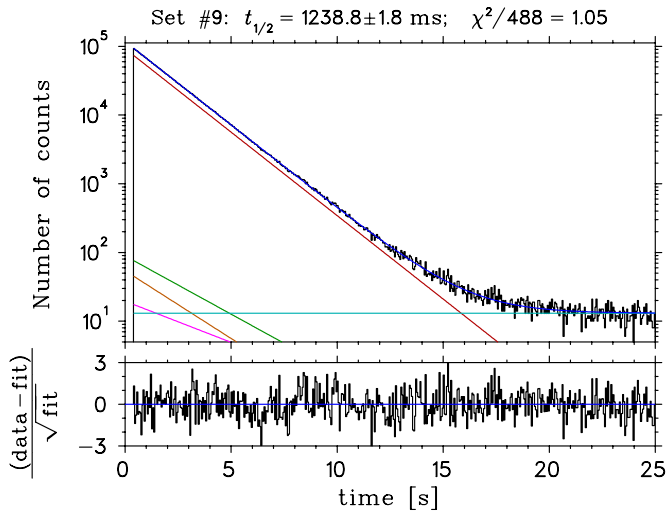}
  \caption{(Color online) Typical summed decay curve of the deadtime-corrected 
    data obtained from a single run (top) and corresponding residuals of the 
    fit (bottom).  The \tsups{37}K decay curve accounts for 99.5\% of the 
    data, with 0.27\% from the \tsups{35}Ar and \tsups{33,34}Cl contaminants, 
    and the rest from background.  The reduced $\chi^2$ of the fit was 1.05 
    with 488 degrees of freedom.\label{fig:ex-summed-fit}}
\end{figure}

\section{Data Analysis}
The selection of data for final analysis was made according to three 
criteria.  First, cycles with too few counts -- less than 5000 -- were 
rejected.  Second, the number of $\beta$'s recorded in each cycle was compared
to the corresponding number of heavy ions recorded in the scintillator at
the exit of MARS.  If their ratio was abnormally low, it meant that the tape 
did not stop with the activity positioned correctly within the $4\pi$ 
proportional gas counter.  In that case, the cycle was rejected. Finally, a 
cycle was rejected if a half-life fit of the recorded data resulted in a very 
poor reduced $\chi^2 >1.3$, which corresponds to a goodness-of-fit statistic 
less than $0.001\%$.

We analyzed the decay data from accepted cycles using two different 
methods: one based on summing all accepted cycles in a run (``summed'' 
analysis) and the other on performing a simultaneous fit on all the cycles of 
a run (``global'' analysis).  All fits employed a fortran code based 
on the Marquardt algorithm for $\chi^2$ minimization.  The 
program assumes Poisson rather than Gaussian statistics in the data.  

The fit function contained a constant background plus four exponentials 
corresponding to the decays of \tsups{37}K and the main contaminants 
\tsups{35}Ar, \tsups{34}Cl and \tsups{33}Cl.  The half-lives of the 
contaminants were fixed at their well-known measured values:  
$t_{1/2}=1.7752(10)~\mathrm{s}$ for \tsups{35}Ar~\cite{SevPRC08}, 
$1.52655(44)~\mathrm{s}$ for \tsups{34}Cl~\cite{Hardy:09} and 
$2.5111(40)~\mathrm{s}$ for \tsups{33}Cl~\cite{SevPRC08}.  The intensity of 
the contaminants could not be fit separately because of the high correlations 
between them, so we fixed the relative normalization of the contaminants 
at the values measured in the MARS focal-plane detector (corrected for losses
in the degraders), and allowed only the total amount of contaminant activity 
to vary relative to \tsups{37}K in the fitting procedure.

\begin{figure}\centering
  \includegraphics[angle=90,width=0.485\textwidth]{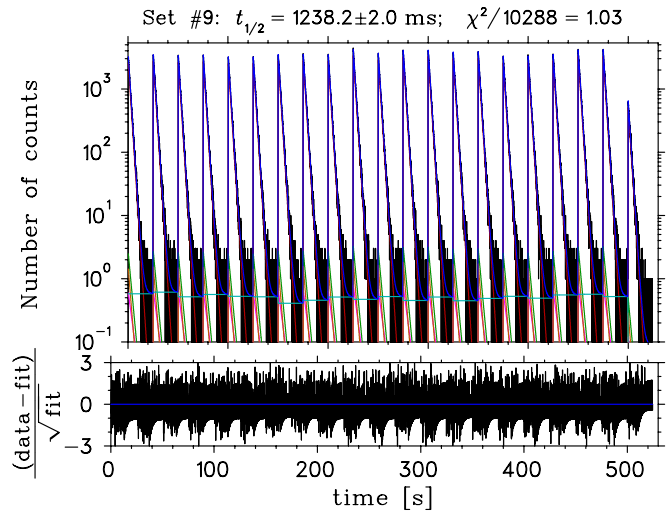}
  \caption{(Color online) Simultaneous fit to the 21 decay curves of a 
    single run containing a total of 311 cycles combined into groups of 15 
    (top) and corresponding residual plot (bottom).  The data here are the 
    same as shown in Fig.~\ref{fig:ex-summed-fit} except in this case the data 
    are not summed to a single decay curve. Here the reduced $\chi^2$ is 1.025 
    with 6858 degrees of freedom.\label{fig:ex-global-fit}}    
\end{figure}

In the ``summed'' method of analysis, the first step was to correct the 
measured decay spectrum in each cycle channel-by-channel for deadtime, based 
on the measured rate in each channel. Next, the cycles of a given run were 
summed into two decay curves, one for each imposed dominant deadtime.  
Finally, these spectra were fitted as already described.  
Figure~\ref{fig:ex-summed-fit} shows a typical deadtime-corrected decay 
curve for a single run analyzed using the summed analysis technique.  Note 
that the fitted background level in the figure agrees well with the value 
obtained in our dedicated background run.

In the ``global'' analysis, the fitting function itself is adjusted to account
for deadtime effects. In this method, many separate decay spectra within a
single run were fit simultaneously, all with the same lifetime for \tsups{37}K but
with separate contamination levels and backgrounds allowed for each 
spectrum. Runs had between 120 and 750 cycles; so, to create a more manageable
number of spectra and also to improve the statistics in the individual spectra,
we summed the cycles together in groups and for each group took the effective
count rate to be the average of the rates in its component cycles. We then
performed a global fit on the resulting summed spectra. The \tsups{37}K half-life 
extracted turned out to be insensitive to the number of cycles summed together 
for each spectrum. Figure~\ref{fig:ex-global-fit} shows a simultaneous 
fit using the global-analysis method; it is of the same run for which the 
summed analysis is illustrated in Fig~\ref{fig:ex-summed-fit}.  The extracted 
values for the half-life obtained from the two analysis methods are well 
within uncertainties, as one can see from the results shown in 
Figs.~\ref{fig:ex-summed-fit} and \ref{fig:ex-global-fit}.

\begin{figure}\centering
  \includegraphics[angle=90,width=0.485\textwidth]{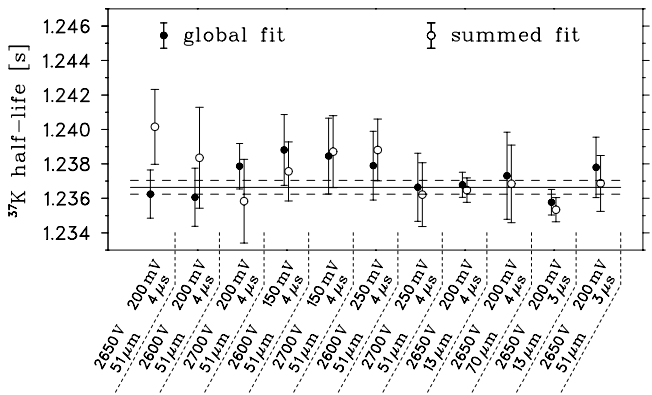}
  \caption{The half-life obtained for \tsups{37}K with different 
    experimental conditions using the two analysis methods.  The labels for
    each run are the detector bias, discriminator setting, degrader thickness 
    and imposed deadtime.  The solid line shows the best-fit half-life with 
    the dashed lines representing the statistical uncertainty in the fit.
    \label{fig:halflife-vs-dataset}}
\end{figure}

\begin{table}[b]
  \caption{Half-life values for \tsups{37}K obtained from two different 
    analysis techniques.  All of the uncertainties shown are purely 
    statistical.  The adopted half-life is taken to be the unweighted 
    average of the two methods.\label{table:I}}
  \begin{ruledtabular}
    \begin{tabular}{lc}
      Method & half-life [s]\\
      \hline
      Summed analysis   & $1.23633(47)$ \\
      Global analysis   & $1.23669(38)$ \\
      \hline
      & \\[-0.9em]
      Average \tsups{37}K half-life  &  $1.23651(47)$  \\[-0.25em]
    \end{tabular}
  \end{ruledtabular}
\end{table}

\subsection{Systematic Uncertainties}
The changing of parameters from run to run allowed us to test 
for potential systematic effects that could contribute to the uncertainty 
in the final results. In Fig.~\ref{fig:halflife-vs-dataset} we plot the 
half-life obtained for 11 different run conditions using both analysis 
techniques. Since two of the 13 runs had the same conditions as another run, 
each has been averaged with its twin and plotted as a single half-life.  Each 
plotted result is labeled by the four measurement parameters that pertained to 
that result: detector bias, discriminator setting, degrader thickness and 
imposed deadtime.  As already mentioned, each run produced two data streams, 
each with a different imposed deadtime.  In all cases, the half-lives obtained 
from the two streams agreed very closely.  Since both contain the same primary 
data, only the result from one of the data streams is presented in Fig.\;3.
No systematic dependence on measurement parameters is evident. 
 
In 
Table~\ref{table:I} we present the half-lives, averaged over all runs, 
obtained from the two different analysis techniques, ``summed''  and 
``global''.  We take our final result to be the unweighted average of these 
two half-lives, with the statistical uncertainty on the average conservatively 
chosen to be the larger of the two individual values.  

\begin{figure}\centering
  \includegraphics[angle=90,width=0.485\textwidth]{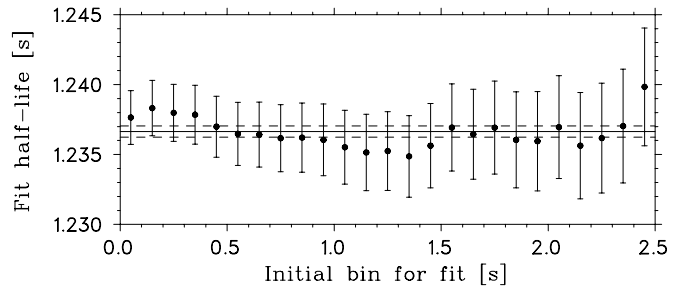}
  \caption{Test for otherwise undetected short-lived impurities or for 
    possible rate-dependent effects. Each point is the result of a separate 
    fit to one of the runs as the starting time of the fitting range is 
    increased.  The solid and dashed lines correspond to the average 
    half-life value and uncertainties given in 
    Fig.~\ref{fig:halflife-vs-dataset}. \label{fig:timethresh}}
\end{figure}

\begin{table}[b]
  \caption{Error budget for the \tsups{37}K lifetime measurement.  All of the 
    values listed are in milliseconds.\label{table:systematics}}
  \begin{ruledtabular}
    \begin{tabular}{lllD{.}{.}{2.2}@{}D{.}{.}{1.2}}
      \multicolumn{3}{l}{Source} & \multicolumn{2}{c}{Uncertainty}\\
      \hline
      \\[-0.8em]
      \multicolumn{3}{l}{Statistical}             && 0.47\\
      \\[-0.8em]
      \multicolumn{3}{l}{Systematics:}\\
      &&Contaminants              & \ 0.72&\\
      &&Analysis method           & \ 0.30&\\
      &&Deadtime                  & \ 0.20&\\ 
      &&Fitting range             & \ 0.13&\\
      &&Bias voltage              & \ 0.04&\\
      &&Discriminator threshold   & \ 0.04&\\
      \cline{4-5}
      \\[-0.8em]
      & \multicolumn{2}{l}{Total systematic}      && 0.82\\  
      \hline
      \\[-0.8em]
      \multicolumn{3}{l}{Total uncertainty}       && 0.94\\[-0.25em]
    \end{tabular}
  \end{ruledtabular}
\end{table}

Although we regularly monitored impurities with the detector at the focal 
plane of MARS and accounted for the observed impurities in our fitting 
procedures, we nevertheless made an additional check at the analysis stage 
for unidentified short-live impurities.  We did this by artificially removing 
up to the first $3.0~\mathrm{s}$ (60~channels) of the counting period in 
steps of $0.1~\mathrm{s}$. At each step the remaining channels in the reduced 
data set were fitted and a half-life extracted. The results are plotted in 
Fig.~\ref{fig:timethresh}, from which it is clear from that the derived 
half-life is stable against these changes.  This test also demonstrates our 
system's independence of counting rate.  We examined the sensitivity of our 
fit results to all the potential sources of systematic errors described above. 
The outcome is the error budget given in Table~\ref{table:systematics}.   We 
found the greatest variation in the extracted lifetime when we fixed the value 
of the total impurity level to the amount observed in the MARS focal-plane 
detector rather than leaving it free to vary in the fits.  The next largest 
source of uncertainty was the counting statistics.  Our final result for the 
half-life of \tsups{37}K is:
\begin{equation}
  1.23651(94)~\mathrm{s}. 
\end{equation}
This $0.08\%$ measurement is nearly an order of magnitude improvement over 
the previously accepted value~\cite{SevPRC08} and now completely dominates 
the world average of the \tsups{37}K half-life.

\section{\boldmath$\mathcal{F}t$-value}
We follow the convention used in Refs.~\cite{NavPRL09} and~\cite{SevPRC08} 
by defining the corrected $\mathcal{F}t$ value for mirror nuclei which 
includes the variation due to the ratio of the Fermi to Gamow-Teller matrix 
elements. Specifically, the $\mathcal{F}t_0$ value is defined as
\begin{align}
  \mathcal{F}t_0 = \mathcal{F}t\ C_V^2\,|M_F^0|^2\,\Big[ 1 + 
  \big(\tfrac{f_{A}}{f_{V}}\big)\rho^{2}\Big]
\end{align}
where experimentally $\mathcal{F}t$ has already been defined in 
Eq.~\eqref{eq:Ft} in terms of the measured $ft$ value and the small 
radiative and isospin-mixing corrections as calculated in 
Ref.~\cite{SevPRC08}. The mixing ratio $\rho$ defined by Eq.~\eqref{eq:rho} 
is experimentally determined from a measurement of one of the angular 
correlations of the decay.
 
When calculating the $\mathcal{F}t$ value, we take the branching ratio and 
total decay energy of \tsups{37}K from the survey of Ref.~\cite{SevPRC08}:  
$\mathrm{BR}=97.99(14)\%$ and $Q_{EC}=6.14746(20)~\mathrm{MeV}$. 
 
Using our new value for the half-life of \tsups{37}K, we find 
\begin{align}
  \mathcal{F}t = 4605.4\pm8.2~\mathrm{s}.
\end{align}
This result is four times more precise than the previous $\mathcal{F}t$ value 
quoted in Ref.~\cite{NavPRL09}, and is now comparable to the precision of the 
three most precisely measured $T=1/2$ mirror nuclei. The branching 
ratio, no longer the half-life, is what now limits the precision of the
$\mathcal{F}t$ value.  We have already made a new measurement of this 
branching ratio and are currently in the process of analyzing the data.  
This should reduce the uncertainty in the $\mathcal{F}t$ value even further.

The only \tsups{37}K correlation measurement published to date, 
$B_\nu=-0.755(23)$~\cite{MelconianPLB2007}, can be used to determine a value 
for $\rho$.  Using the recoil-order corrections of Holstein~\cite{holstein}, 
we find $\rho=-0.560^{+24}_{-29}$, where the asymmetric error bars arise from 
the fact that the uncertainty in $B_\nu$ is so large that the uncertainty on 
the extracted value of $\rho$ is not Gaussian.  Using this value of $\rho$ and 
the $\mathcal{F}t$ value above leads to
\begin{equation}
 \mathcal{F}t_{0} = 6057^{+149}_{-122}~\mathrm{s}.
\end{equation}
With our new lifetime measurement, the central value of $\mathcal{F}t_{0}$ 
has changed and is now better aligned with other mirror transitions. 

\begin{figure}[t]
\centering
  \includegraphics[angle=90,width=0.485\textwidth]{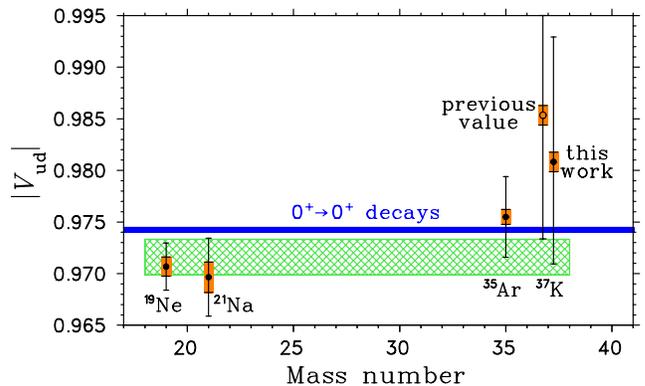}
  \caption{(Colour online) Values of $V_{ud}$ deduced for the four most 
    precisely measured mirror transitions as a function of mass number. The 
    smaller error bars with the filled in region represent the component of 
    the total uncertainty arising from just $\mathcal{F}t$ (the total error 
    bars are all dominated by the correlation measurements).  The hatched 
    region represents the fit to the average ($\chi^2/3=0.7$) yielding 
    $|V_{ud}|=0.9718(17)$.  For comparison, the solid band shows the value 
    of $V_{ud}$ derived from the pure Fermi transitions.\label{fig:Vud-summary}}
\end{figure}

Finally, noting that for a mixed transition, $\mathcal{F}t_0$ is related 
to $V_{ud}$ by~\cite{NavPRL09}
\begin{align}
 V^2_{ud} = \frac{K}{\mathcal{F}t_{0}\,G_{F}^{2}\,(1+\Delta_{V}^{R})},
\end{align}
we calculate that from \tsups{37}K alone
\begin{equation}
  |V_{ud}|= 0.981^{+12}_{-10}.
\end{equation}
This can be compared to the older value of 0.985(12). The uncertainty is 
barely affected by our improved half-life measurement because it is 
dominated by the large uncertainty in the $B_\nu$ correlation parameter 
measurement.  

In Fig.~\ref{fig:Vud-summary} we plot $V_{ud}$ values for the four most 
precisely measured nuclear mirror transitions. The point for \tsups{19}Ne has 
been updated from the survey of Ref.~\cite{SevPRC08} to include recent lifetime 
results~\cite{triambakPRL2012,ujicPRL2013,broussardPRL2014}.  The figure shows 
that the main effect of the present \tsups{37}K lifetime measurement is to 
shift its value of $V_{ud}$ into better agreement with the other mirror nuclei 
and the $0^+\!\!\rightarrow0^+$transitions.

\section{Conclusion}
We have measured the half-life of \tsups{37}K, the decay of which proceeds 
primarily via a mirror $3/2^+\rightarrow3/2^+$ $\beta^+$ transition to its 
analog, the ground state in \tsups{37}Ar.  Our new result for the half-life 
is $1.23651(94)~\mathrm{s}$. The $0.08\%$ precision of this result is nearly 
an order of magnitude improvement over the previously accepted world average.  
The $\mathcal{F}t$ value for the \tsups{37}K mirror transition is now 
determined to $0.17\%$ precision, and the corresponding values of 
$\mathcal{F}t_0$  and $V_{ud}$ have shifted into better alignment with values 
determined for the other well-measured mirror transitions.  Nevertheless, the 
uncertainties of $\mathcal{F}t_0$  and $V_{ud}$ remain completely dominated by 
the correlation measurement for the transition.  The results from improved 
correlation measurements, currently in progress by the \trinat{} collaboration 
at \triumf\;\cite{melconianLASNPA}, are required 
before the uncertainty on the mirror-transition value for $V_{ud}$ can be 
improved.

\section{Acknowledgements}
We are grateful to the support staff of the Cyclotron Institute, especially 
Don May and George Kim for providing the primary beam.  This work was supported by the U.S. 
Department of Energy under Grant No.\ DE-FG02-93ER40773 and Early Career Award 
ER41747, as well as the Robert A. Welch Foundation under Grant No. A-1397.

\bibliography{k37-lifetime-PRC-refs}

\end{document}